\documentclass[amsmath,amssymb,prl,twocolumn]{revtex4-1}
\usepackage{dcolumn}
\pdfoutput=1
\usepackage[english]{babel}
\usepackage[utf8]{inputenc}
\usepackage[T1]{fontenc}
\usepackage{amsthm}
\usepackage{amsfonts}
\usepackage{url}
\usepackage{tensor}
\usepackage[colorlinks = true,
            linkcolor = blue, 
            linkcolor = blue,
            urlcolor  = blue,
            citecolor = blue,
            anchorcolor = blue]{hyperref}
\usepackage{graphicx}
\usepackage{mathrsfs}
\usepackage{enumitem}
\usepackage{caption}
\usepackage[usenames,dvipsnames]{xcolor}

\newcounter{mnotecount}[section]

\newcommand{\beq}{\begin{eqnarray}}
\newcommand{\eeq}{\end{eqnarray}}
\newcommand{\ben}{\begin{eqnarray*}}
\newcommand{\een}{\end{eqnarray*}}

\newtheorem*{theorem*}{Theorem}
\theoremstyle{definition}

\newcommand\ringring[1]{%
  {
   \mathop{\kern0pt #1}\limits^{
     \vbox to-1.85ex{
       \kern-2ex 
       \hbox to 0pt{\hss\normalfont\kern.1em \r{}\kern-.45em \r{}\hss}
       \vss 
     }   }
  }
}

\begin{document}
\title{Chaos and Einstein--Rosen gravitational waves}
\author{Sebastian J. Szybka}
\author{Syed U. Naqvi}
\affiliation{Astronomical Observatory, Jagiellonian University}
\begin{abstract}
We demonstrate the existence of chaotic geodesics for the Einstein--Rosen standing gravitational waves. The complex dynamics of massive test particles are governed by a chaotic heteroclinic network. We present the fractal associated with the system under investigation. Gravitational standing waves produce intricate patterns through test particles in a vague analogy to mechanical vibrations generating Chladni figures and complicated shapes of Faraday waves.
\end{abstract}
\maketitle{}

\section{Introduction}

Simple physical settings can give rise to complex phenomena --- a fact that has been recognized in many cultures long before the discovery of deterministic chaos. The seminal example of such processes is provided by the interplay between standing mechanical waves and material properties, as observed by Leonardo da Vinci and others \cite{wl1982,chladni,faraday}. In this paper, we demonstrate that standing gravitational waves are no different from mechanical waves in this respect. Vibrations of a spacetime induce complex behavior of test particles, which, in the system studied, is manifested through chaos.

The standing gravitational waves, studied in this paper, belong to a class of cylindrically symmetric time-dependent vacuum solutions to the Einstein equations known as the Einstein-Rosen waves. Formally, these spacetimes were discovered by Beck \cite{Beck25} in $1925$. However, their importance was not recognized until they were rediscovered by Einstein and Rosen in $1937$ \cite{EinsteinRosen37}. These solutions played a prominent role in the history of the gravitational waves \cite{bwaves} confirming clearly, for the first time beyond the linear regime, that general relativity predicts existence of gravitational waves. Einstein and Rosen showed that each solution of the cylindrical wave equation in a flat spacetime leads via quadratures to the solution of the vacuum Einstein equations --- a cylindrical gravitational wave. After the seminal work of Einstein and Rosen, the cylindrical gravitational waves have been studied by Marder \cite{marder}, Thorne \cite{thorne}, Stachel \cite{Stachel1966}, Chandrasekhar \cite{chandra}, and others. (For a more complete bibliography see the books \cite{exact2003,est}.) These waves provided a precious insight into the nonlinear nature of gravitation \cite{PiranSafier:1985}.

In their original paper, Einstein and Rosen mention, among other solutions, stationary waves (standing waves \cite{Stephani:2003,Hermann:2004,Szybka:2019}). In this manuscript, we show the existence of chaotic geodesics for a particular simple member of this class (a subclass of solutions investigated in \cite{Rosen54,Halilsoy:1988,Stephani:2003,ers,Szybka:2019}). We demonstrate, for the first time, that standing gravitational waves provide a natural setting for chaotic behavior. (Chaotic geodesics and spacetimes have been extensively studied in general relativity, e.g., in the articles \cite{PhysRevD.98.084050,podolsky98,podovesely,Szybka:2007fx,Conto1,Conto2}.)

Test particles in standing wave spacetimes have been investigated beyond cylindrical symmetry. In the models studied \cite{SzybkaNaqvi:2021,hal}, the amplitude of waves diminished as space expanded. Consequently, the influence of waves on test particles was transient and intricate patterns have not been observed.

Finally, we note that an experimental setup for generating weak, nearly cylindrical standing gravitational waves was proposed a long time ago \cite{GrishchukSazhin1975,RevModPhys.52.341}, but it still remains beyond our current technical capabilities. 

\section{Setting}

The cylindrically symmetric metric with two hypersurface orthogonal Killing fields $\partial_{z}$, $\partial_\varphi$ can be written in geometrized units in the form
\begin{equation}\label{metric}
g=e^{2(\gamma-\psi)}\left(-dt^2+d\rho^2\right)+\rho^2e^{-2\psi}d\varphi^2+e^{2\psi}dz^2\;,
\end{equation}
where $\rho>0$, $-\infty<t,z<\infty$, $0\leq\varphi<2\pi$ and the metric functions $\psi$ and $\gamma$ depend on $t$ and $\rho$ only. The Einstein equations take a well-known form
\begin{equation}
\psi''+\frac{1}{\rho}\psi'-\ringring{\psi}=0\;,\;\;\mathring{\gamma}=2\rho\mathring{\psi}\psi'\;,\;\;\gamma'=\rho\left(\mathring{\psi}^2+\psi'^2\right)\,,\nonumber
\end{equation}
where circles and primes denote $\partial_t$, $\partial_\rho$, respectively.
Hereafter, we consider the simplest standing cylindrical wave 
\begin{equation}\label{af}
\begin{split}
	\psi&=\frac{1}{2}J_0\!\!\left(\frac{\rho}{\lambda}\right)\cos\!\left(\frac{t}{\lambda}\right)\;,\\
	\gamma&=\frac{\rho^2}{8\lambda^2}\left[ J_0^2\!\!\left(\frac{\rho}{\lambda}\right)+J_1^2\!\!\left(\frac{\rho}{\lambda}\right)
	-2\frac{\lambda}{\rho} J_0\!\!\left(\frac{\rho}{\lambda}\right)J_1\!\!\left(\frac{\rho}{\lambda}\right)\cos^2\!\!\left(\frac{t}{\lambda}\right)\right]\!,\nonumber
\end{split}
\end{equation}
where $\lambda$ is a constant and $J_i$ are the Bessel functions of the first kind and $i$th order. The metric \eqref{metric} with the auxiliary functions defined above constitutes the regular exact solution to the vacuum Einstein equations that will be studied in this paper. (The Weber-Wheeler \cite{RevModPhys.29.509}, Bonnor \cite{bpulse} pulse is a related famous solution.)

Let $u=u^\alpha \partial_\alpha=\dot{t}\partial_t+\dot{\rho}\partial_\rho+\dot{\varphi}\partial_\varphi+\dot{z}\partial_z$ denote the four-velocity of a massive test particle parametrized by the proper time $\tau$ (the dot denotes differentiation in $\tau$). We have $g(u,u)=-1$. The symmetries of the spacetime imply that $$g(\partial_z,u)=u_z=e^{2\psi}\dot{z}\;,\;\;\; g(\partial_\varphi,u)=u_\varphi=\rho^2e^{-2\psi}\dot{\varphi}=l$$ are conserved along the geodesics. The constant $l$ plays a role of an angular momentum per unit mass (without a loss of generality we may assume $l\geq 0$). For simplicity, we consider timelike geodesics restricted to the plane $z=const$ which implies $\dot{z}=u^z=u_z=0$. The geodesic equation can be reduced to the first-order system
\begin{equation}
\begin{split}
	\dot{t}&=T\;,\\	\dot{T}&=-(T^2+P^2)\Delta_{,t}-2 T P \Delta_{,\rho}+e^{-2\Delta+2\psi}\psi_{,t}\, \frac{l^2}{\rho^2}\;,\\
	\dot{\rho}&=P\;,\\
	\dot{P}&=-(T^2+P^2)\Delta_{,\rho}-2 T P \Delta_{,t}-e^{-2\Delta+2\psi}(\psi_{,\rho}-\frac{1}{\rho})\, \frac{l^2}{\rho^2} \;,\nonumber
\end{split}
\end{equation}
where $\Delta=\gamma-\psi$ is a new auxiliary function and $\varphi(\tau)$ may be found via quadratures $\varphi=\int l/\rho^2 e^{2\psi}d\tau$ once the system above is solved.  The equations above contain an apparent singularity (for $\rho=0$ the coordinates are not well-defined). This singularity is irrelevant for our studies because for $l=0$ the singular terms vanish and for $l\neq 0$ the conservation of angular momentum prevents particles from reaching $\rho=0$. The explicit form of the system above is too large to be usefully presented here. The first integral $g(u,u)=-1$ may be used for simplification, but we find it more convenient not to do so and keep it as a constraint which provides a consistency check. All numerical calculations presented in this article were performed for $l=m/3$, $\lambda=m/(2\pi)$. We relied on Wolfram Mathematica for initial computations, while more demanding tasks were handled in C++. We also used independent Python code to cross-verify most of our results. The absolute and relative tolerances in the C++ ODE solver  were set to $10^{-6}$ and $10^{-10}$, respectively. We employed the additive Runge-Kutta method from the SUNDIALS library \cite{sundials1}. In Wolfram Mathematica's NDSolve procedure, we used the default tolerances and allowed for an automatic method selection. 

The geodesics in a similar spacetime (the cylindrical standing waves coupled to a massless scalar field) and in its high frequency limit have been studied elsewhere \cite{cieslik}, but no glimpse of chaos has been observed at that time.

\onecolumngrid
\begin{widetext}
\begin{figure}[h!]
\includegraphics[width=0.45\linewidth]{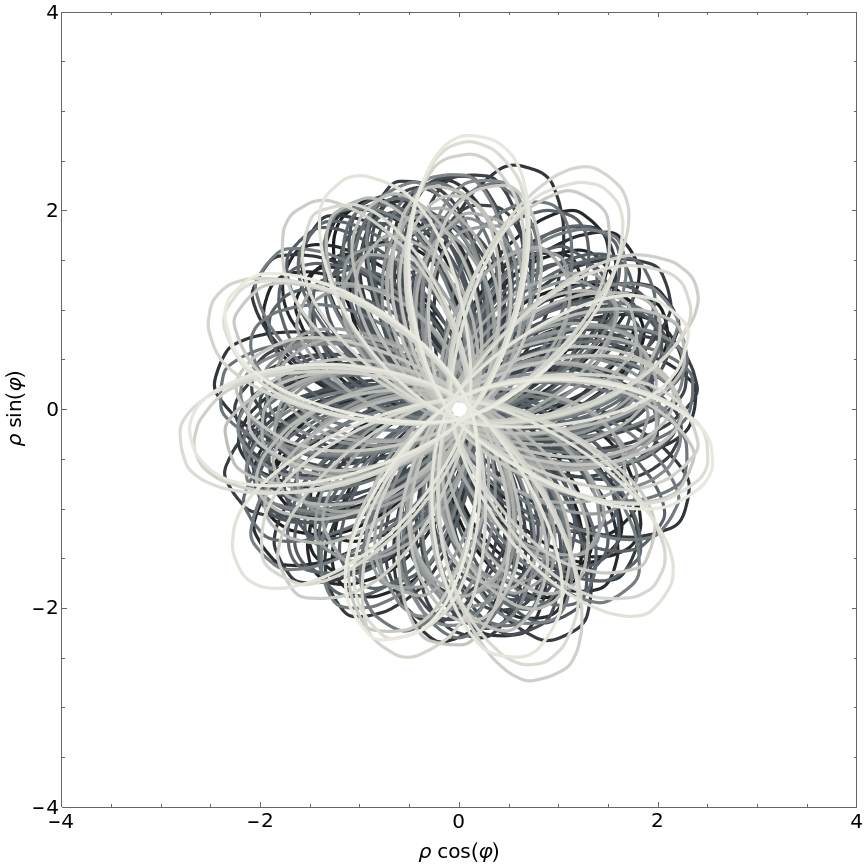}
\hfill
\includegraphics[width=0.45\linewidth]{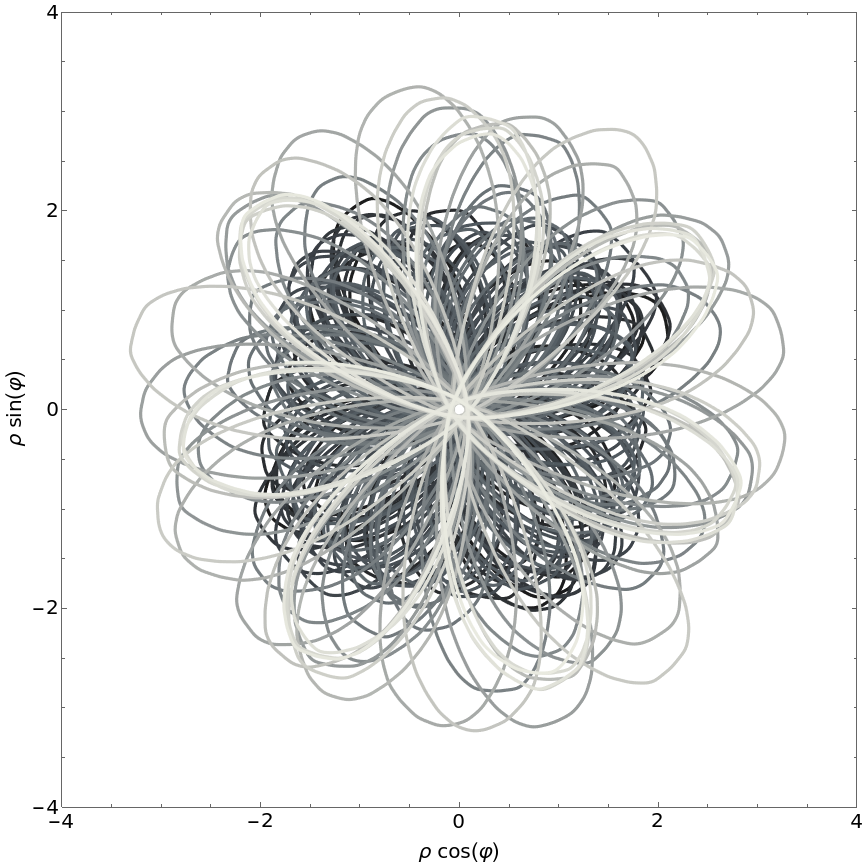}
	\caption{Two timelike geodesics starting at slightly different initial points differ dramatically revealing sensitivity to initial conditions. The left figure corresponds to $t_0=0$, $\rho_0=1.13062m$, $P_0=0.83928$, $\tau_0=0$, $\varphi_0=0$, and $\tau_{max}=3000m$ [the value of $T_0>0$ follows from the first integral $g(u,u)=-1$]. The right figure corresponds to the same parameters and initial conditions with $P_0$ shifted by $+6\times 10^{-5}$. The darker shades of gray correspond to initial part of the trajectories.}
\label{fig1}
\end{figure}
\end{widetext}

\twocolumngrid

\section{Geodesics}

The cylindrical symmetry implies that the curved cylindrically symmetric spacetime cannot be asymptotically flat. Moreover, for the spacetime studied, the $(2+1)$-dimensional Lorentzian submanifolds $z=const$ with a naturally induced metric are not asymptotically flat. The metric function $\psi$ is bounded in the limit $\rho\rightarrow +\infty$, but $\gamma$ blows up to $+\infty$ (which follows from the asymptotic behavior of the Bessel functions). The blow up of $\gamma$ implies that $t=const$ Riemannian metrics are not asymptotically flat \cite{afg2}. This observation is consistent with the notion of C-energy introduced by Thorne in \cite{thorne} (the C-energy equals $\gamma/4$). 

The first integrals $g(u,u)=-1$ and $\dot\varphi=e^{2\psi}l/\rho^2$ can be combined into the equation
\begin{equation}\label{constr}
-1=e^{2(\gamma-\psi)}(-\dot{t}^2+\dot{\rho}^2)+e^{2\psi}\frac{l^2}{\rho^2}\;,
\end{equation}
which supplemented with the asymptotic behavior of the metric functions implies that all timelike geodesics are bounded in $\rho$. The timelike geodesics with $l\neq 0$ circle the symmetry center. The amplitude of the standing wave is largest at the symmetry axis and decreases \mbox{radially with $\rho^{-2}$.} Far away from the center, the gravitational field resembles the gravitational field of an infinite strut. Near the center, the test particles' orbits are visibly perturbed by the oscillations of the standing gravitational waves. These waves play the role of the periodic driving ``force.'' In such a setting, the chaos is expected.

Indeed, the numerical studies of the phase space indicate that nonchaotic regions are interwound with the chaotic ones. Two exemplary trajectories presented in Fig.\ \ref{fig1} show sensitivity to initial conditions. The shapes of these trajectories depend on the precision used by the numerical solver; however, the shadowing theorem \cite{ott} implies that errorless trajectories with slightly different initial conditions remain close to the ones presented.

\onecolumngrid
\begin{widetext}
\begin{figure}[h!]
\includegraphics[width=\linewidth,angle=0]{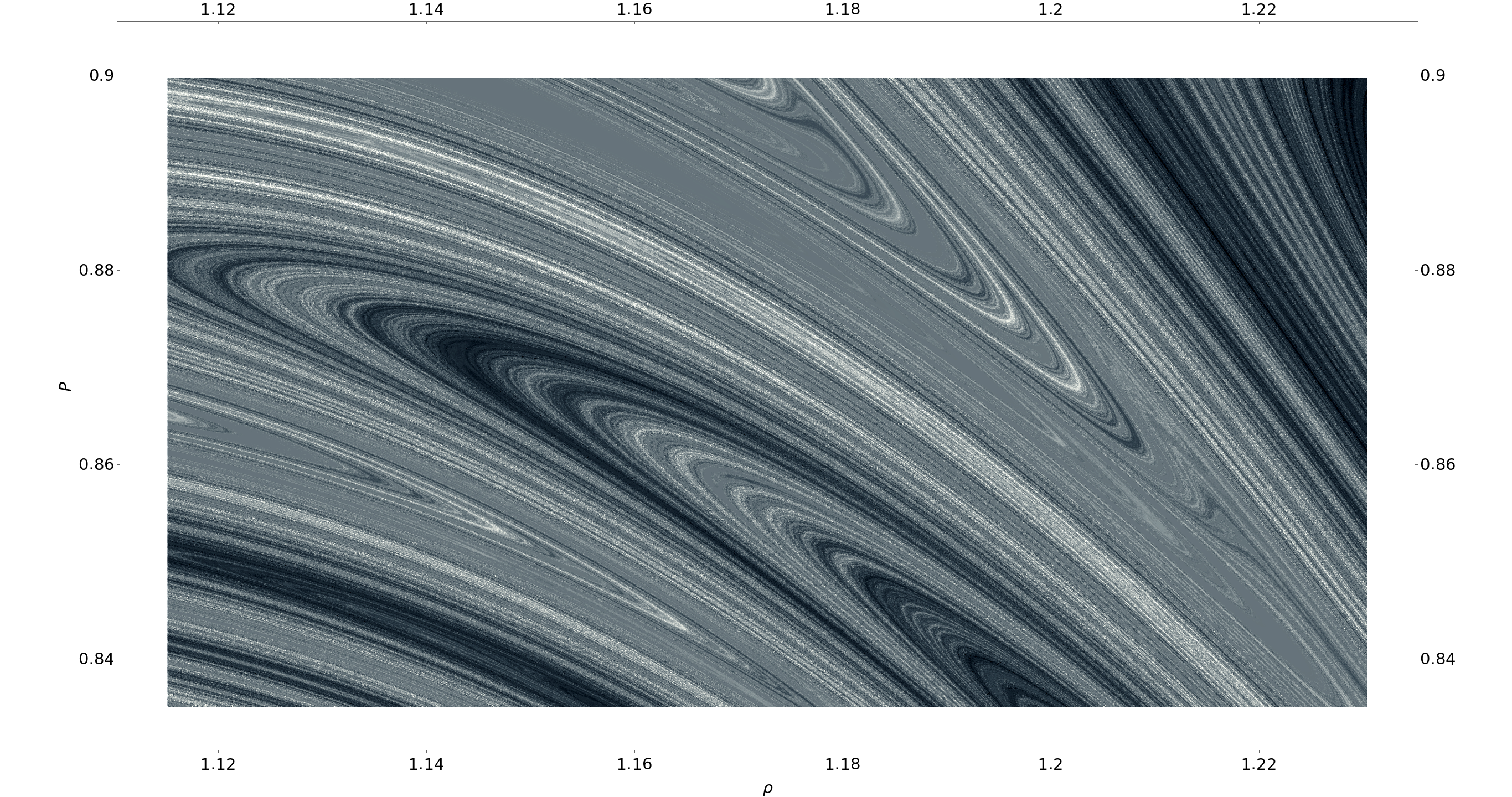}
	\caption{The value of $\varphi(\tau=400m)$ as a function of initial points for $\varphi_0=0$. The darkest point corresponds to $\varphi(400m)=85.87$, the brightest to $\varphi(400m)=247.66$. The winding numbers for these points are $13$ and $39$, respectively. The fractal structure is evident.}
\label{fig2}
\end{figure}
\end{widetext}

\twocolumngrid

\section{Fractal}

The numerical evidence for the sensitivity to initial conditions and chaos can be strengthened by systematic studies of the phase space. In Fig.\ \ref{fig2} we show how $\varphi(\tau=400m)$ varies with initial conditions (its numerical value has been indicated by the fading shades of gray). The fractal structure is clearly visible. 

The Fig.\ \ref{fig2} reveals that the sensitivity to initial conditions presented in Fig.\ \ref{fig1} is not accidental, but represents typical behavior of the system in the large part of the phase space. Some regions in the phase space (not visible in Fig.\ \ref{fig2}) are not chaotic. For example, far away from the symmetry center standing waves are not strong enough to substantially perturb orbits of test particles to induce a chaotic behavior.

\section{Chaotic heteroclinic network}

In the previous sections, the numerical evidence for chaos has been presented. In this section, we apply the method of Poincaré sections to reveal a mechanism that induces complicated dynamics of the system. 

\begin{widetext}
\begin{figure}[h!]
\begin{minipage}[t]{0.4\linewidth}
\includegraphics[width=\linewidth,angle=0]{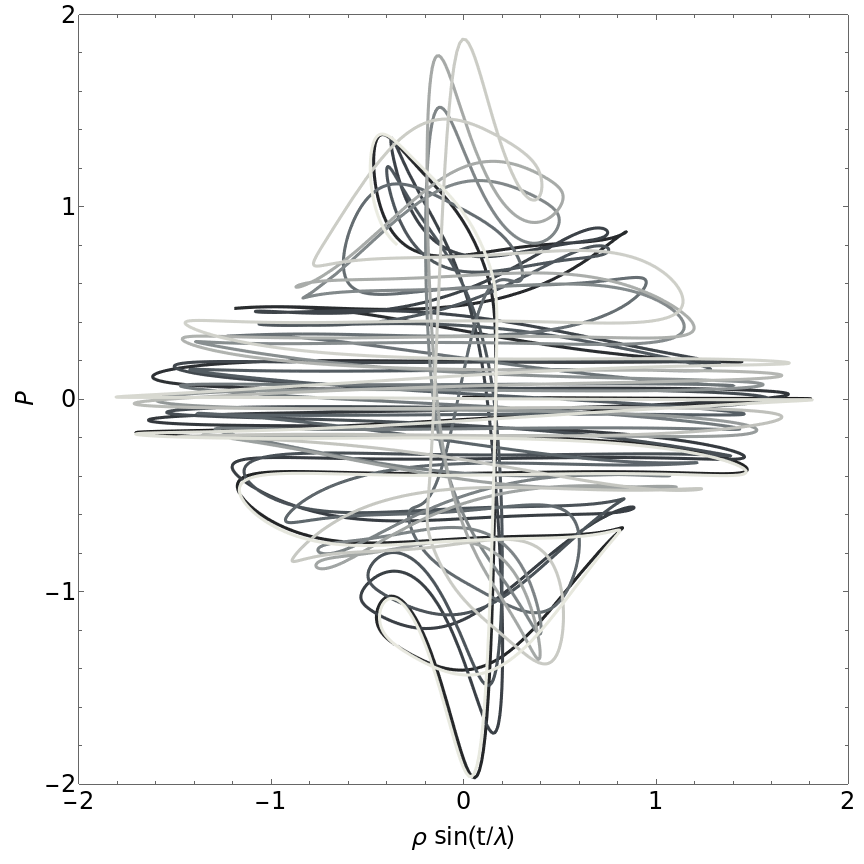}
	\caption{The projection of the trajectory in the phase space in which $t/ \lambda$ is an angular variable. The trajectory starts at the point $t_0=0$, $\rho_0=1.8$, $P_0=0.01$. The proper time $\tau$ varies from $0$ to $63$. The darker shades of gray corresponds to the initial part of the trajectory.}
\label{fig3}
\end{minipage}
\begin{minipage}[t]{0.4\linewidth}
	\includegraphics[width=0.545\linewidth,angle=0]{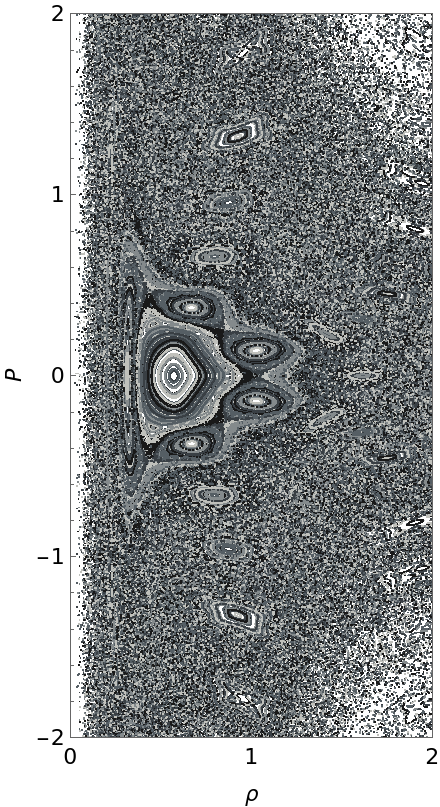}
	\caption{The Poincaré section of the trajectories along $t/\lambda = 2\pi k$, where $k=0,1,2,\dots$. We eliminate $T=\dot{t}$ from the geodesic equation using the constraint \eqref{constr}. The shades of gray represent different trajectories. The periodic and quasiperiodic orbits are immersed in a sea of chaotic ones.}
\label{fig4}
\end{minipage}
\end{figure}
\end{widetext}

The geodesic equation depends on the coordinate $t$ only through the term $\cos(t/\lambda)$ and its derivatives. Therefore, it is convenient to treat $t/\lambda$ as an angle. The projection of a trajectory in the phase space is presented in Fig.\ \ref{fig3}.

In order to understand the dynamics of the system it is convenient to reduce the three-dimensional flow to a two-dimensional Poincaré map $\phi$  with a help of the Poincaré section $t\; mod\;2\pi\lambda = 0$ (see Fig.\ \ref{fig4}). 

\onecolumngrid

\twocolumngrid

The Poincaré section reveals existence of numerous periodic points surrounded by Kolmogorov-Arnold-Moser tori (KAM tori) \cite{ott}. The KAM tori correspond to quasiperiodic orbits. The periodic points are fixed points or exhibit $5$-periodicity. Hereafter, for the sake of clarity, we consider $\phi^5$ instead of $\phi$, hence all $5$-periodic points become fixed points of the $\phi^5$ Poincaré map. In the central portion of the section presented in Fig.\ \ref{fig4}, one of the fixed points is encircled by five elliptic fixed points. One of these elliptic fixed point is barely visible and lies on the axis $P=0$, within the KAM tori elongated along the $P$ axis. These fixed points arose from annihilated resonant tori and, in accordance with the Poincaré-Birkhoff theorem \cite{birkhoff1913}, are accompanied by five hyperbolic fixed points $p_{i=0,\dots,4}$ located at the junctions of the KAM tori. We show below that these five hyperbolic fixed points are important for the behavior of the system in the central part of the phase space.

\begin{figure}[h!]
        \includegraphics[width=0.745\linewidth,angle=0]{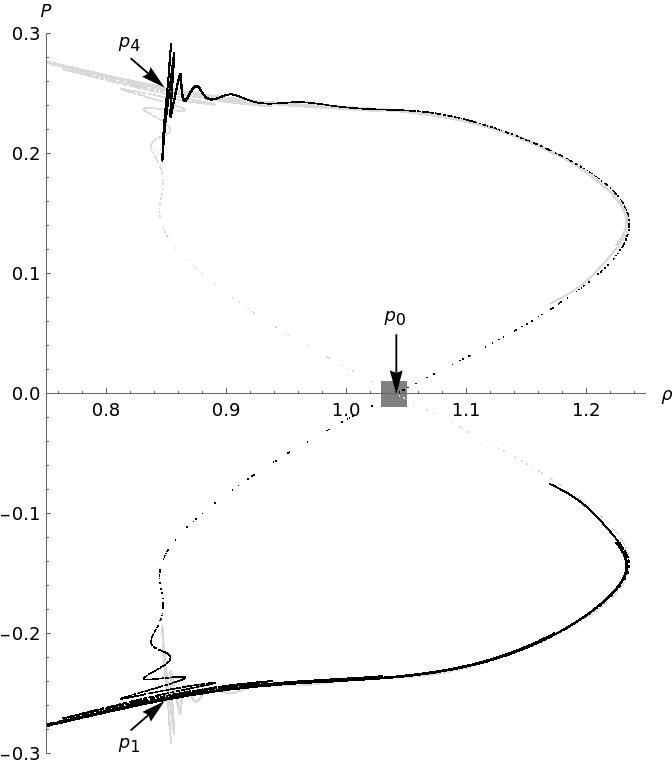}
	\caption{The stable and unstable manifolds of the hyperbolic fixed point $p_0$ cross, forming two homoclinic tangles near $p_1$ (bottom) and $p_4$ (top), namely $S(p_0)\cap U(p_0)\neq \varnothing$. The darker color corresponds to the unstable manifold $U(p_0)$. The set $R$ (the shaded rectangle) contains $p_0$.}
\label{fig5}
\end{figure}

\begin{figure}[h!]
        \includegraphics[width=0.845\linewidth,angle=0]{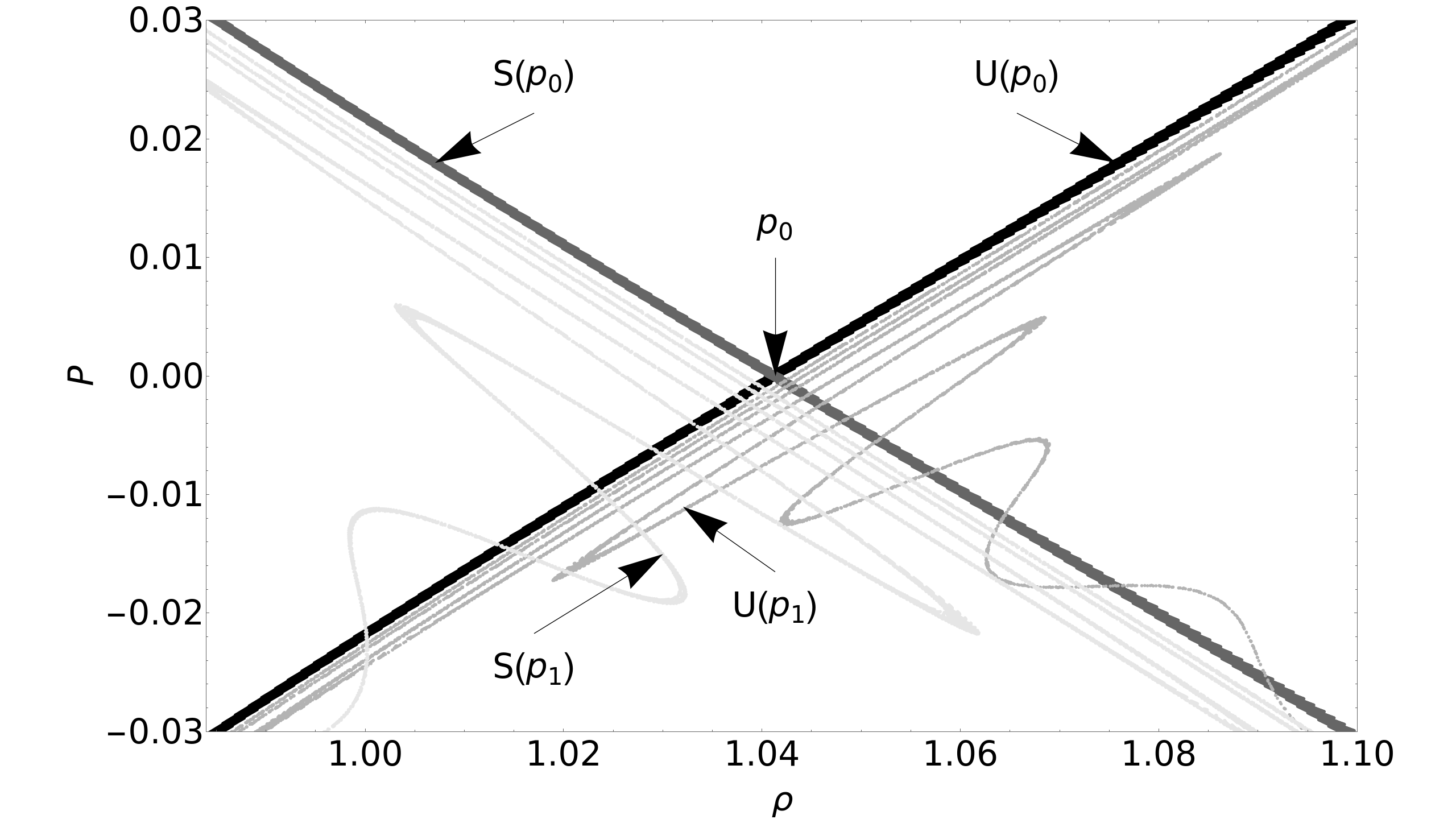}
	\caption{The stable and unstable manifolds of the hyperbolic fixed points $p_0$ and $p_1$ cross forming heteroclinic intersections near $p_0$, namely $S(p_0)\cap U(p_1)\neq\varnothing\land U(p_0)\cap S(p_1)\neq\varnothing$. The homoclinic intersection is also visible ($S(p_1)\cap U(p_1)\neq\varnothing$). The manifolds are marked by shades of gray. 
}
\label{fig6}
\end{figure}

\begin{figure}[h!]
        \includegraphics[width=0.745\linewidth,angle=0]{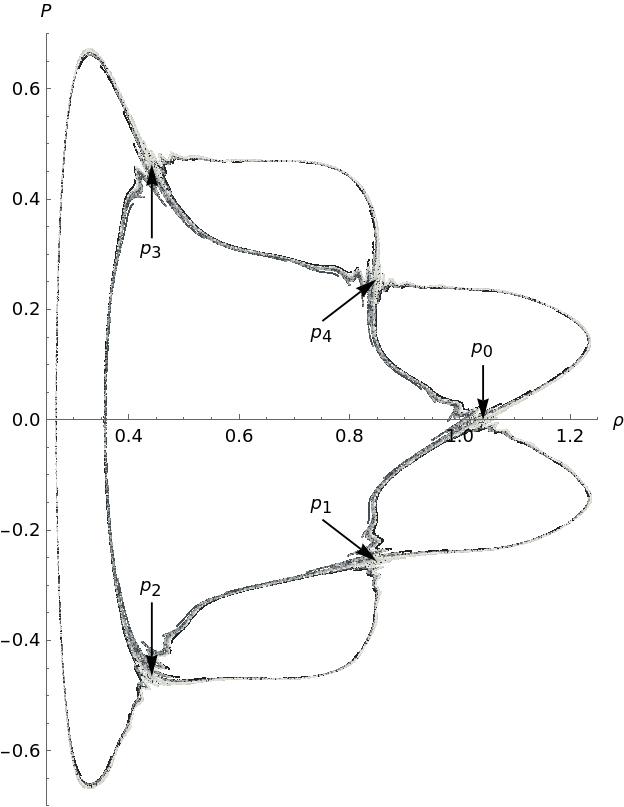}
        \caption{The chaotic heteroclinic network revealed by forward and backward iterations of the set $R$ with $\phi$. The shades of gray indicate different trajectories. }
\label{fig7}
\end{figure}

\begin{figure}[h!]
        \includegraphics[width=0.745\linewidth,angle=0]{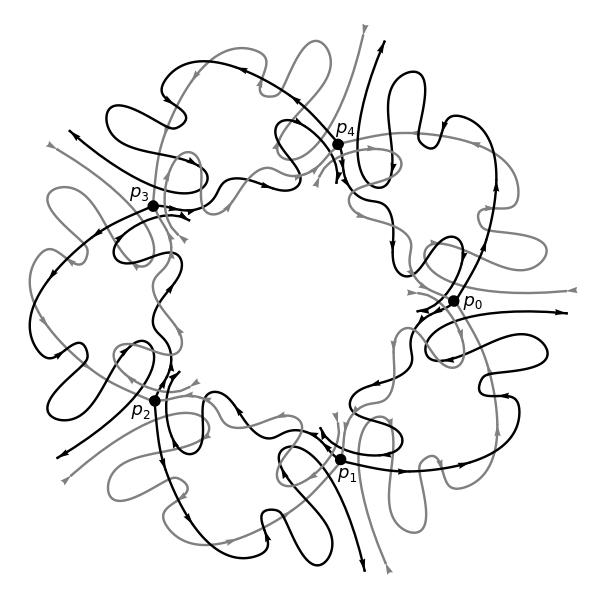}
        \caption{The symbolic representation of the chaotic heteroclinic network. The darker color corresponds to the unstable manifolds.}
\label{fig8}
\end{figure}

Let $S(p_i)$ and $U(p_i)$ denote stable and unstable manifolds of the hyperbolic fixed point $p_i$. We consider one of the five hyperbolic points, namely $p_0\simeq(1.041334,0)$, a small real constant $\epsilon$ and an open set $R$ such that $p_0\in R$ and for every $p'\in R$ we have $d(p',p)<\epsilon$, where $d$ corresponds to the Euclidean distance in the phase space. The subsequent iterations of sets $\phi^{-5}(R)$ and $\phi^{5}(R)$ reveal shapes of the stable and unstable manifolds of $p_0$. We demonstrate in Fig.\ \ref{fig5} that the stable manifold and unstable manifold of $p_0$ cross ($S(p_0)\cap U(p_0)\neq \varnothing$) leading to the Poincaré homoclinic tangle which is a well-known phenomenon implying the existence of chaotic trajectories. Homoclinic intersections of stable and unstable manifolds are accompanied by heteroclinic intersections of stable and unstable manifolds of different hyperbolic fixed points. In Fig.\ \ref{fig6}, we present heteroclinic intersections of stable and unstable manifolds which belongs to $p_0$ and $p_1=(0.848293,-0.255512)$ [$S(p_0)\cap U(p_1)\neq\varnothing\land U(p_0)\cap S(p_1)\neq\varnothing$].  Similar structures are formed by stable and unstable manifolds of all hyperbolic fixed points $p_i$. They create a five-component chaotic heteroclinic network, which is revealed through subsequent iterations of $\phi(R)$ and $\phi^{-1}(R)$ in Fig.\ \ref{fig7}. The chaotic heteroclinic network is depicted symbolically in Fig.\ \ref{fig8}.

\clearpage

\newpage

\twocolumngrid

\section{Summary}

We presented numerical evidence for the existence of chaotic geodesics in the cylindrical standing gravitational wave spacetime of the Einstein--Rosen form. The phenomenon was explained using the method of Poincaré sections which revealed that the system's dynamics are driven by a chaotic heteroclinic network. The complex effect of the standing gravitational wave on test particles is analogous to the well-known effect of standing mechanical waves in Newtonian physics.

Finally, we note the interesting formal coincidence of the mathematical structures: the heteroclinic networks and chaotic heteroclinic networks are critical systems with a wide range of applications in the modeling of cognitive processes and neural dynamics \cite{TCD,chaoticHN}. These networks offer computational capabilities, known as heteroclinic computing, and can process information similarly to how neural networks operate \cite{dcHN,PhysRevLett.109.018701}. 

\vspace{0.2cm}

\noindent{\sc Acknowledgments}

We thank Julia Osęka and Kamil Wyrwich for comments.
This research was supported in part by PLGrid Infrastructure (Grant No.\ PLG/2022/015759).
We acknowledge the contributions of OpenAI's GPT-4 to the production of Fig.\ \ref{fig8} and the resolution of linguistic nuances. The calculations were conducted in C++ and Python with the SUNDIALS library \cite{sundials1,sundials2,bjorndahlgren}, and in Wolfram Mathematica with the xAct package \cite{xAct}.

\bibliographystyle{apsrev4-1}
\setcitestyle{authortitle}
\bibliography{report}

\end{document}